\documentclass[final,5p,times,twocolumn,fleqn]{elsarticle}

\usepackage{geometry}
\usepackage{amsmath,bm,upgreek}

\biboptions{numbers,sort&compress}

\usepackage{lipsum}
\usepackage{flushend}
\usepackage{float}
\usepackage{siunitx}
\usepackage{amsfonts}
\usepackage{enumitem}
\usepackage[tableposition=top]{caption}
\usepackage{subfig}
\usepackage{makecell}
\usepackage{xurl}
\usepackage[usenames,dvipsnames]{color}
\usepackage{gensymb}
\usepackage{booktabs}

\usepackage{caption}
\usepackage{subcaption}

\usepackage{threeparttable}

\usepackage{listings}
\usepackage{xcolor}

\definecolor{codegreen}{rgb}{0,0.6,0}
\definecolor{codegray}{rgb}{0.5,0.5,0.5}
\definecolor{codepurple}{rgb}{0.58,0,0.82}
\definecolor{backcolour}{rgb}{0.95,0.95,0.92}

\lstdefinestyle{mystyle}{
    backgroundcolor=\color{backcolour},   
    commentstyle=\color{codegreen},
    keywordstyle=\color{magenta},
    numberstyle=\tiny\color{codegray},
    stringstyle=\color{codepurple},
    basicstyle=\ttfamily\footnotesize,
    breakatwhitespace=false,         
    breaklines=true,                 
    captionpos=b,                    
    keepspaces=true,                 
    showspaces=false,                
    showstringspaces=false,
    showtabs=false,                  
    tabsize=2
}

\usepackage{mathtools}
\DeclarePairedDelimiter\abs{\lvert}{\rvert}%
\makeatletter
\let\oldabs\abs
\def\abs{\@ifstar{\oldabs}{\oldabs*}}
\makeatother

\newcommand{\Cs}[2]{#1_\text{#2}}
\newcommand{\dd}[1]{\mathrm{d}#1}

\begin{document}
\sisetup{separate-uncertainty=true}

\begin{frontmatter}
\author[soheil]{Soheil Solhjoo\corref{cor1}}
\cortext[cor1]{soheil@solhjoo.com; s.solhjoo@rug.nl}
\affiliation[soheil]{organization={ENTEG, Faculty of Science and Engineering, University of Groningen},
            addressline={Nijenborgh 4}, 
            city={Groningen},
            postcode={9747 AG}, 
            country={the Netherlands}}

\title{Flow curve approximation using Avitzur’s model
for barreling compression test}

\begin{abstract}
The combination of the Cylindrical Profile Model (CPM) and Avitzur's model is commonly used to determine flow stress curves in material testing using compression test. In this process, stress is corrected for friction using Avitzur's model and the average strain is calculated from CPM. This study proposes a method for estimating strain based on Avitzur's model. The presented case studies demonstrate the impact of this strain correction on the flow curves. The results show that increasing friction leads to higher strain values at the center of the sample due to barreling. The proposed method provides a more accurate interpretation of compression test results.
\end{abstract}

\end{frontmatter}


\section*{Introduction}\label{s:introduction}
The Cylindrical Profile Model (CPM) and Avitzur's model \cite{Avitzur1969} are often employed together to determine flow stress curves of materials. This process involves compressing the material at a specific and constant strain rate $\dot{\varepsilon}$. According to CPM, this can be achieved by controlling the ramp velocity $U$ as a function of the sample's instantaneous height $H$:
\begin{equation}
    U = H \dot{\varepsilon}
    \label{eq:CPM_U}
\end{equation}
The corresponding strain can then be calculated using:
\begin{equation}
    \Cs{\varepsilon}{CPM} = \ln \left( H / H_0 \right)
    \label{eq:CPM_strain}
\end{equation}
where $H_0$ denotes the sample's initial height.
While it is straightforward to establish the relationship between the recorded compression force $F$ and the average stress $\overline{\sigma}$ using CPM, this model does not account for barreling. To address this limitation, Avitzur's model is commonly employed instead \cite{khoddam2021state}.

Avitzur proposed a simplified approximation that makes certain assumptions; see, e.g., \cite{Avitzur1969, solhjoo2019evaluation} for detailed information. Avitzur's model describes the material flow within the sample of an axial compression test (ACT) using a 2D velocity field, which results in a detailed relationship between the compression force and the average stress, a function of friction factor $m$ unknown in experimental tests.
Recently, Solhjoo introduced a combination of Avitzur's model with CPM and presented A-CPM as follows \cite{solhjoo2023actas}:
\begin{equation}
    \frac{F/S}{\overline{\sigma}} = 1 + \left( \frac{\overline{R}}{H} \right)^2 / \left( \frac{6}{b}-1 \right)
    \label{eq:A-CPM}
\end{equation}
where 
$\overline{R}$ is the effective radius, and $b$ is an arbitrary coefficient representing the deformed sample's barreling that can be obtained as a function of the sample's geometry. The results obtained from A-CPM are nearly identical to the solutions provided by Avitzur's model. Still, it benefits from a concise form and is not directly dependent on the unknown parameter $m$.

Although Avitzur's model is often utilized for determining deformation stress, strain is typically calculated using CPM. This work proposes an alternative solution for estimating the sample's strain and strain rate based on Avitzur's model.

\section*{Strain field of Avitzur's velocity field}

The strain rate components of Avitzur's model are obtained to be:
\begin{subequations}
    \begin{flalign}
        \Cs{\dot{\varepsilon}}{r} &= \dot{\varepsilon}_\uptheta = \frac{AU}{H} e^{-bz/H} \\
        \Cs{\dot{\varepsilon}}{z} &= - 2 \Cs{\dot{\varepsilon}}{r} \\
        \Cs{\dot{\varepsilon}}{rz} &= - \frac{br}{2H} \Cs{\dot{\varepsilon}}{r} \\
        \dot{\varepsilon}_{\text{r}\uptheta} &= \dot{\varepsilon}_{\text{z}\uptheta} = 0 
    \end{flalign}
\end{subequations}
Here, $U$ is the velocity of the compressing ramp, and
\begin{equation}
    A = \frac{b/4}{1 - e^{-b/2}}
\end{equation}
Using these components, the effective strain rate is given by:
\begin{equation}
    \dot{\varepsilon} = \Cs{\dot{\varepsilon}}{r} \sqrt{4+\frac{1}{3} \left( \frac{br}{2H} \right)^2} \approx 2 \Cs{\dot{\varepsilon}}{r}
    \label{eq:eff_strain_rate}
\end{equation}
The approximation is valid because Avitzur's model assumes $b^n=0$ for $n \geq 2$.

With the strain rate $\dot{\varepsilon}$ available, calculating the strain $\varepsilon = \int \dot{\varepsilon} \dd{t}$ is straightforward. However, Avitzur's model provides a strain and strain rate distribution throughout the sample. To examine these properties at the center of the sample ($z=0$), we use the following equations:
\begin{subequations}
    \begin{flalign}
        \dot{\varepsilon} &= 2 \frac{AU}{H} \label{eq:Avitzur_strate} \\
        \Cs{\varepsilon}{Avitzur} &= \int \dot{\varepsilon} dt = 2A \int_0^t \frac{U}{H}\dd{t} 
        \label{eq:Avitzur_strain_1}
    \end{flalign}
\end{subequations}
By considering $U \dd{t} = -\dd{H}$, equation \ref{eq:Avitzur_strain_1} is rewritten as follows.
\begin{equation}
    \begin{split}
        \Cs{\varepsilon}{Avitzur} &= 2A \int_H^{H_0} \frac{\dd{H}}{H} = 2A \ln \left( H / H_0 \right) \\
        &= 2A \Cs{\varepsilon}{CPM}
        \label{eq:Avitzur_strain}
    \end{split}
\end{equation}
This equation establishes the relationship between the effective strains based on CPM and Avitzur's model. In other words, when calculating the stress values using Avitzur's model or A-CPM (equation \ref{eq:A-CPM}), the corresponding strain should be obtained from equation \ref{eq:Avitzur_strain}.

Furthermore, it is important to note that controlling the ramp velocity according to CPM (equation \ref{eq:CPM_U}) does not ensure a constant strain rate according to Avitzur's model. To achieve a constant strain rate using Avitzur's model, one can use equation \ref{eq:Avitzur_strate} to determine a ramp velocity of $U = 0.5 H\dot{\varepsilon} / A$. However, it should be emphasized that $A$ is a function of $b$, which has no unique and reliable solution. All suggested solutions of $b$ are functions of sample geometry, including top- and mid-plane radii in addition to height \cite{solhjoo2019evaluation}.

\section*{Case study}
A series of compression tests are performed to demonstrate the impact of this correction on the flow curve of a sample. A cylindrical sample with an initial radius of $R_0=5 \si{mm}$ is compressed from an initial height of $H_0=16 \si{mm}$ to a final height of $\Cs{H}{f}=6 \si{mm}$ at a constant strain rate of $1 \si{\per\second}$. To maintain this constant strain rate, the ramp speed is controlled using equation \ref{eq:CPM_U} according to CPM.

The test is simulated using the ACTS module of a recently introduced model \cite{solhjoo2023actas}. The mechanical behavior of the material is defined by the relationship $\sigma = 100 \varepsilon^{0.2} \dot{\varepsilon}^{0.01} \si{\mega\pascal}$, and various friction factors of $m = \{0.01, 0.3, 0.5, 0.7, 1\}$ are investigated. It should be noted that $m=0$ is not examined in this study, as it results in a barreling parameter of $b=0$, leading to an indeterminate parameter $A$.

\section*{Results and discussion}
All samples exhibited the foldover phenomenon, making them incompatible with Avitzur's model. However, recent studies have shown that Avitzur's model can still approximate the flow stress curves of such samples \cite{solhjoo2023actas}. The following is an attempt to interpret test results using  Avitzur's model.

It is well known that there is no unique method to accurately estimate the barreling parameter of Avitzur's model \cite{solhjoo2019evaluation}. To interpret the test results, I examined various available solutions and found that the most accurate estimates are obtained using the following equation:
\begin{equation}
    b = 2\frac{\Delta R}{\Delta H} \left( \frac{H_0}{R_0} + \frac{1}{2} \left( \frac{H}{\overline{R}} + \frac{\Cs{H}{f}}{\Cs{\overline{R}}{f}} \right)\right)
    \label{eq:Avitzur_b_full}
\end{equation}
where $\Delta H = H_0 - H$ represents the change in height, and $\Delta R = \Cs{R}{M} - \Cs{R}{T}$ corresponds to the difference between the mid-plane radius $\Cs{R}{M}$ and the top-plane radius $\Cs{R}{T}$. This equation represents a weighted average of equations A.1 and A.4 in \cite{solhjoo2019evaluation}, obtained from a static framework. It is important to note that this calculation should not incorporate instantaneous measurements unless $\Delta R$ is available at any instantaneous $H$, which is not possible by the currently available test rigs \cite{khoddam2021state}; instead, the equation utilizes data from only the initial and final states to obtain a single parameter value, reducing equation \ref{eq:Avitzur_b_full} to the mean of equations A.1 and A.4 in \cite{solhjoo2019evaluation} as follows, which is used for the further calculations in this investigation.
\begin{equation}
    b = 2\frac{\Delta R}{\Delta H} \left( \frac{H_0}{R_0} + \frac{\Cs{H}{f}}{\Cs{\overline{R}}{f}} \right)
    \label{eq:Avitzur_b}
\end{equation}

\begin{figure*}[t]
  \centering
    \includegraphics[page=1,width=0.33\textwidth]{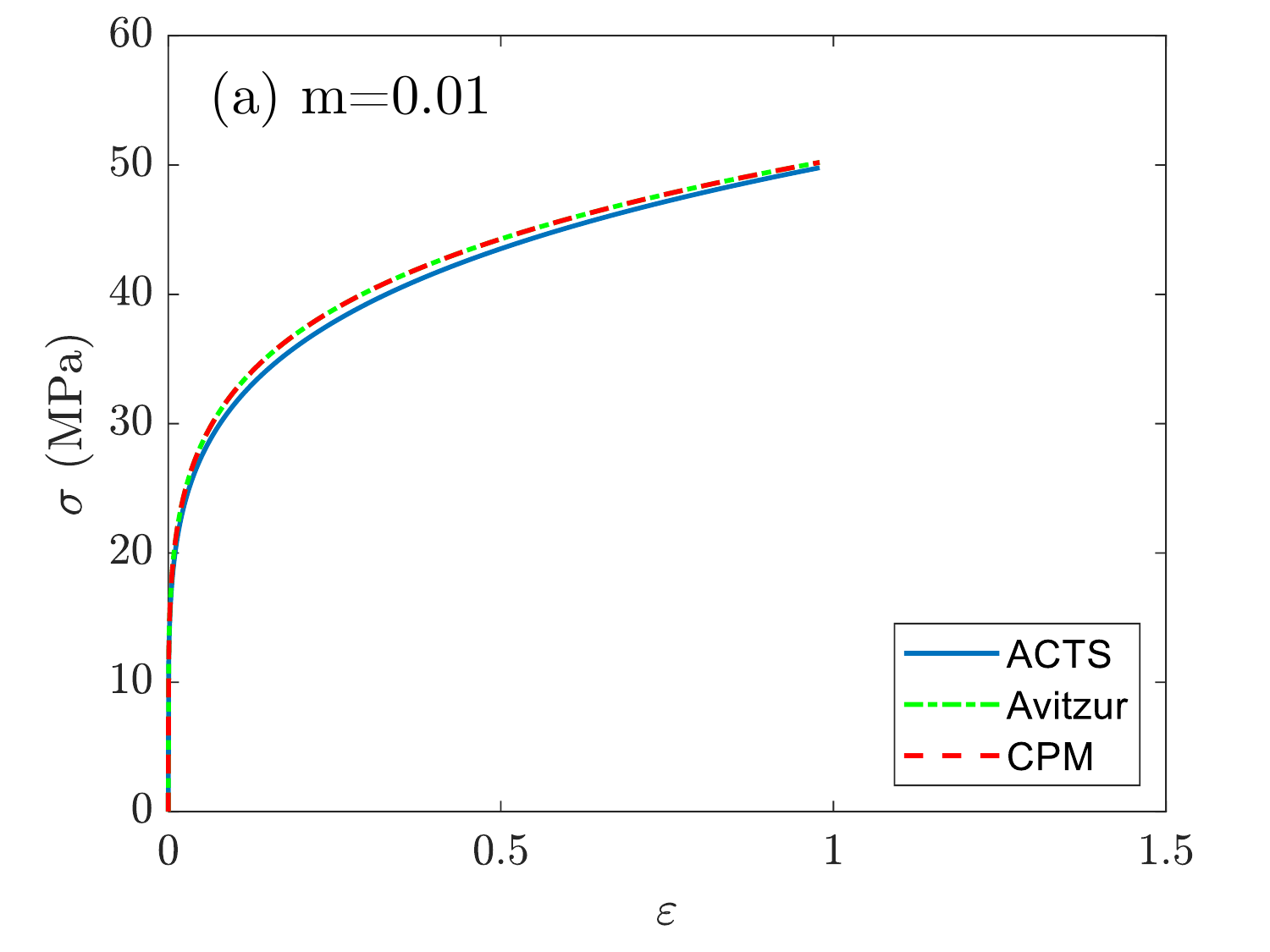}
    \includegraphics[page=2,width=0.33\textwidth]{figs.pdf}
    \includegraphics[page=3,width=0.33\textwidth]{figs.pdf}
    \includegraphics[page=4,width=0.33\textwidth]{figs.pdf}
    \includegraphics[page=5,width=0.33\textwidth]{figs.pdf}
  \caption{The stress-strain curves at the center of the sample with friction of (\textbf{a}) 0.01, (\textbf{b}) 0.3, (\textbf{c}) 0.5, (\textbf{d}) 0.7, and (\textbf{e}) 1. The difference between the results of CPM (equation \ref{eq:CPM_strain}) and Avitzur's model (equation \ref{eq:Avitzur_strain}) is only for their strain values.}
  \label{fig:ss_curves}
\end{figure*}

The results, presented in figure \ref{fig:ss_curves}, show that the solutions of CPM and Avitzur's model for $m=0$ are identical, indicating that $\lim_{m \to 0} A = 0.5$. 
Samples exhibit more pronounced barreling as friction increases and experience higher strain throughout their volume. The change in barreling behavior is captured by the parameter $A$, enabling a more accurate interpretation of the compression test. In contrast, the strain obtained from CPM remains constant at $\Cs{\varepsilon}{CPM}=\ln (16/6) \approx 1$.

When friction is increased to $m=1$, the approximated stresses deviate significantly from the expected solutions and can falsely suggest that the stress has reached its maximum. This discrepancy arises from the inaccurate estimation of $b$, reported in the literature \cite{solhjoo2019evaluation, solhjoo2023actas, khoddam2022critical}.

\section*{Summary and conclusions}
In this study, Avitzur's model for the disk compression test is investigated, and a formula for calculating strain at the center of the sample is derived. By utilizing the recorded force-displacement data and measured sample geometry at its initial and final stages, the flow stress curve at the center of the sample can be estimated from the following:

\begin{subequations}
    \begin{flalign}
        \frac{F}{\overline{\sigma}} &= \pi \overline{R}^2 \left( 1 + \left( \frac{\overline{R}}{H} \right)^2 / \left( \frac{6}{b}-1 \right) \right)\\
        \varepsilon &= 2A \ln \left( H / H_0 \right)       
    \end{flalign}
\end{subequations}
with
\begin{subequations}
    \begin{flalign}
        \overline{R} &= R_0 \sqrt{H_0 / H} \\
        b &= 2\frac{\Delta R}{\Delta H} \left( \frac{H_0}{R_0} + \frac{\Cs{H}{f}}{\Cs{\overline{R}}{f}} \right) \\
        A &= \frac{b/4}{1 - e^{-b/2}}        
    \end{flalign}
\end{subequations}
where $b$ takes data from only the initial and final stages; however, if $\Delta R$ is available corresponding to all recorded heights, $b$ can be estimated using equation \ref{eq:Avitzur_b_full} for higher accuracy.

The proposed model demonstrates good performances for friction factors $m \leq 0.7$. However, it is essential to emphasize that the friction factor is generally unknown in experimental setups and requires determination through relevant and dependable theories. The findings presented in this study align with existing literature, indicating that Avitzur's model lacks overall reliability. This limitation has prompted the development of new theories aimed at providing accurate interpretations of compression test \cite{khoddam2021state,solhjoo2019evaluation, solhjoo2023actas, khoddam2022critical, khoddam2021verified, khoddam2019power}.

\appendix

\section{Accurate effective strain rate}
Assuming $b^n=0$ for $n \geq 2$, which is valid in Avitzur's model, leads to $\dot{\varepsilon} \approx 2 \Cs{\dot{\varepsilon}}{r}$ (see equation \ref{eq:eff_strain_rate}); however, one can calculate the effective strain rate without taking the approximation as follows:

\begin{subequations}
    \begin{flalign}
        \dot{\varepsilon} &= A (G_0 + \Cs{G}{f}) \label{eq:eff_strain_rate_full} \\
        \Cs{G}{i} &= \ln \left( \frac{\Cs{C}{i} + 2 \Cs{H}{i}}{\Cs{C}{i} - 2 \Cs{H}{i}} \right) - \Cs{C}{i}/\Cs{H}{i} \\
        \Cs{C}{i} &= \sqrt{4 \Cs{H}{i}^2 - B} \\
        B &= \frac{1}{3} \left( \frac{br}{2} \right)^2.
    \end{flalign}
\end{subequations}
The discrepancies between the two solutions (equations \ref{eq:eff_strain_rate} and \ref{eq:eff_strain_rate_full}) are negligible.

\bibliographystyle{elsarticle-num}
\bibliography{cas-refs}
\end{document}